\documentclass[]{spie}  
\usepackage[]{graphicx}

\usepackage{amsmath}
\usepackage{amssymb}

\title{What shakes the FX tree? Understanding currency dominance, dependence and dynamics}

\author{Neil F. Johnson\supit{a}, Mark McDonald\supit{b}, Omer Suleman\supit{a}, Stacy Williams\supit{c} and  Sam Howison\supit{b}
\skiplinehalf
\supit{a}Physics Department, Oxford University, Oxford OX1 3PU, U.K. \\
\supit{b}Mathematics Department, Oxford University, Oxford, OX1 2EL, U.K. \\
\supit{c}FX Research and Trading Group, HSBC Bank, 8 Canada Square, London E14 5HQ,
U.K.
}

\authorinfo{Further author information: (Send correspondence to N.F.J.)  \\
E-mail: n.johnson@physics.ox.ac.uk, Telephone: +44 1865 272287}

\begin{document} 
  \maketitle

\newcommand{\av}[1]{\left\langle#1\right\rangle}
\newcommand{\£}{\pounds}
\newcommand{\E}{\mathrm{E}}
\newcommand{\Var}{\mathrm{Var}}
\newcommand{\Cov}{\mathrm{Cov}}
\newcommand{\x}{\underline{x}}

\begin{abstract}
There is intense interest in understanding the stochastic and dynamical properties of the global Foreign Exchange (FX) market, whose daily transactions exceed $10^{12}$ US dollars. This is a formidable task since the FX market is characterized by a web of fluctuating exchange rates, with subtle inter-dependencies which may change in time. In practice, traders talk of particular currencies being `in play' during a particular period of time -- yet there is no established machinery for detecting such important information. Here we apply  the construction of Minimum Spanning Trees (MSTs) to the FX market, and show that the MST can capture important features  of the global FX dynamics. Moreover, we show that the MST can help identify momentarily dominant and dependent currencies.  
\end{abstract}


\keywords{Fluctuations, correlations, networks, econophysics, graphs}

\section{INTRODUCTION}
\label{sect:intro}  

The interdisciplinary field of complex systems and networks, and in particular the study of their associated dynamical and stochastic properties, is growing rapidly \cite{econo,nets1,nets2,DM2003}. Example applications, include biology, sociology, and even economics and finance through the so-called field of Econophysics \cite{econo,book}. Financial markets offer the possibility of studying a system of great practical importance, but also one where large amounts of accurate data are now available. Indeed it is true that throughout history, the latest technologies have always been employed in order to maximize the accuracy of the recorded prices. The Foreign Exchange (FX) market is arguably the most important of all markets, because of its truly global nature and the fact that it is in continual operation -- it simply never closes. It is also the largest market in the world, with a daily transaction total which exceeds the Gross Domestic Product (GDP) of most countries \cite{taylor}. An understanding of how inter-connected the various currencies are, and how this is reflected in the country-country exchange rates, is therefore of great academic and practical interest \cite{taylor,PEV2003}. Yet despite its importance, the FX market is still relatively poorly understood -- for example, the recent fall in the value of the dollar against other
major currencies is quite mysterious and has attracted numerous economic `explanations' to reason away its dramatic decline. 

Here we present an analysis of a correlation network which characterizes the fluctuating exchange-rates of the major currencies in the FX market. The technical approach which we adopt, is motivated by recent research within the Econophysics community by Mantegna and others \cite{MRN1999,MS2000,JPO2002,OCKK2003b,OCKK2003a,OCKK2003c,GB2003,GB2004}. In particular, we focus on the construction and interpretation of  Minimum Spanning Trees (MST) which are special types of network in which there are relatively few connections and yet no network node remains unconnected. Mantegna and co-workers focused mainly on equities -- by contrast, we consider the case of FX markets and focus on  what the time-dependent properties of the MST can tell us about the FX market's evolution. In particular, we investigate the stability and time-dependence of the resulting MST, and introduce a methodology for inferring which currencies are `in play' by analyzing the clustering and leadership structure within the MST network. 

The application of MST analysis to financial stock (i.e. equities) was introduced by the physicist Rosario Mantegna \cite{MRN1999}. The MST gives a `snapshot' of such a system; however, it is the temporal evolution of such systems, and hence the evolution of the MSTs themselves, which motivates our research. In a series of papers \cite{OCKK2003b,OCKK2003a,OCKK2003c}, Onnela et al. extended Mantegna's work to investigate how such trees evolve over time in equity markets. Here we follow a similar approach for FX markets.  One area of particular interest in FX trading is to identify which (if any) of the currencies  are `in play' during a given period of time. More precisely, we are interested in understanding whether particular currencies appear to be assuming a dominant or dependent role within the network, and how this changes over time. Since exchange rates are always quoted in terms of the price of one currency compared to another, this is a highly non-trivial task. For example, is an increase of the value of the euro versus the dollar primarily because of an increase in the intrinsic value of euro, or a decrease in the intrinsic value of the dollar, or both? We analyze FX correlation networks in an attempt to address such questions. We believe that our findings, while directly relevant to FX markets, could also be relevant to other complex systems containing $n$ stochastic processes whose interactions evolve over time.

\section{NETWORKS, TREES AND THE MST}
\label{sect:MST}  
We begin with a brief review of the properties of networks. A typical network or `graph' contains $n$ nodes or `vertices' $\{i\}$ connected by $M$ connections or `edges'. In the case of a real physical connection such as a road or a wire, it is relatively easy to assign a binary digit (i.e. 1 or 0) to the edge between any two nodes $i$ and $j$  according to whether the corresponding physical connection exists or not. However, for correlation networks of financial securities the identification of network connections is less clear. In fact it is extremely difficult to assign any particular edge as being a definite zero or one -- instead, all edges will typically carry a
weighting value $\rho_{ij}$ which is analog rather than binary, and which is in general neither
equal to zero nor to one. The analysis of such weighted networks is in its infancy, in particular with respect to their functional properties and dynamical evolution \cite{econo}.  The main difficulty is that the resulting network is fully-connected with $M=n(n-1)/2$ connections between all $n$ nodes (since $\rho_{ij}\equiv \rho_{ji}$). For any reasonable number of nodes, the number of connections is very large (e.g.  for $n=110$, $n(n-1)/2=5995$) and hence it is extremely difficult to deduce which correlations are most important for controlling the overall dynamics of the system. Indeed, it would be highly desirable to have a simple method for deducing whether certain nodes, and hence a given subset of these stochastic processes, are actually `controlling' the correlation structure \cite{note9}. In the context of FX trading, such nodal control would support the popular notion among traders that certain currencies can be `in play'
over a given time period. Clearly such information could have important practical consequences in terms of understanding the overall dynamics of the highly-connected FX market. It could also have practical applications in other areas where $n$ inter-correlated stochastic process are operating in parallel. 

Starting with a given correlation matrix (e.g. of financial returns) a connected graph can be constructed by means of a transformation between correlations and suitably defined distances \cite{MS2000}. This transformation assigns smaller distances to larger correlations \cite{MS2000}. The MST contains $n-1$ connections which connect together all $n$ nodes, hence classifying it as belonging to the subset of networks known as trees. It can be constructed from the resulting hierarchical graph \cite{RTV1986,MS2000}. Consider $n$ different time-series labelled by $i$, where $i\in \{1,2,...n\}$. Each time-series can be represented as a vector ${\underline x}_i$ with $p$ components corresponding to the $p$ timesteps, each denoted as $x_{ik}$ where $k\in \{1,2,...p\}$ and where $p \in \mathbb{N}^{+}$ is the same for each timeseries. The corresponding $n \times n$ correlation matrix $C$ is easy to construct, and has elements $C_{ij}\equiv\rho_{ij}$ where 
\begin{eqnarray} \rho_{ij} & = & \frac{ \langle {\underline x}_i . {\underline x}_j \rangle - \langle {\underline x}_i \rangle \langle {\underline x}_j \rangle}{\sigma_i \sigma_j}
\end{eqnarray} 
where $\langle$...$\rangle$ indicates a time-average over the $p$ datapoints labelled by $k\in \{1,2,...p\}$, and
$\sigma_i$ is the sample standard deviation of the time-series ${\underline x}_i$. From the form of $\rho_{ij}$ it is obvious that $C$ is a symmetric matrix. In addition, 
\begin{eqnarray} 
\rho_{ii} = \frac{\langle {\underline x}_i . {\underline x}_i \rangle - \langle {\underline x}_i \rangle ^2}
{\sigma_i^2} & \equiv & 1,  \ \ \forall\  i
\end{eqnarray} 
hence all the diagonal elements are identically 1. Therefore $C$ has $n(n-1)/2$ independent elements. Since the number of relevant correlation coefficients increases like $n^2$, even a relatively small number of time-series can yield a correlation matrix which contains an enormous amount of information -- arguably `too much' information for practical
purposes. By comparison, the MST provides a skeletal structure with only $n-1$ links, and hence attempts to strip the system's complexity down to its bare essentials. As shown by Mantegna, the practical justification for using the MST lies in its ability to provide economically meaningful information \cite{MRN1999,MS2000}. Since the MST contains only a subset of the information from the correlation matrix, it cannot tell us anything which we could not (in principle) obtain by analyzing the matrix $C$ itself. However, as with all statistical tools, the hope is that it can provide an insight into the system's overall behavior which would not be
so readily obtained from the (large) correlation matrix itself.

In order to build the MST, we first need to convert the correlation matrix $C$ into a `distance' matrix $D$. We use the non-linear mapping \cite{MRN1999,MS2000}:
\begin{eqnarray} 
d_{ij}(\rho_{ij}) &=& \sqrt{2(1-\rho_{ij})}
\end{eqnarray} 
to get the elements $d_{ij}$ of $D$ \cite{note1}. Since $-1\leq \rho_{ij} \leq 1$, we have
$0\leq d_{ij} \leq 2$. In particular:
\begin{eqnarray*}
\rho_{ij} = -1 & \longmapsto & d_{ij} = 2 \\
\rho_{ij} = 0  & \longmapsto & d_{ij} = \sqrt{2}\\
\rho_{ij} = \frac{1}{2} & \longmapsto & d_{ij} = 1\\
\rho_{ij} = 1  & \longmapsto & d_{ij} = 0
\end{eqnarray*}
This distance matrix $D$ can be thought of as representing a fully connected graph with edge weights $d_{ij}$. In the terminology of graph theory, a forest is a graph where there are no cycles \cite{BBo1979} while a tree is a connected forest. Thus a tree containing $n$ nodes must contain precisely $n-1$ edges \cite{DM2003,BBo1979}. The minimum spanning tree $\bf{T}$ of a graph is the tree containing every node, such that the sum $\sum_{d_{ij}\in \bf{T}}{d_{ij}}$ is a minimum. There are two methods for constructing the MST --- Kruskal's algorithm and Prim's algorithm \cite{JPO2002}. We use Kruskal's algorithm \cite{CLR1990}.

\section{CLUSTER ANALYSIS}
\label{sect:clust}
As stated earlier, the impetus for this research came from the MST work of Mantegna and colleagues in the Econophysics community. However the task of building a hierarchical clustering corresponding to a particular set of timeseries, actually falls firmly within the established field of cluster analysis. There are two crucial steps in cluster analysis\cite{PH2003}: first one needs to define a measure of the proximity of two timeseries (the `dissimilarity measure'). Then one needs to specify a clustering technique. Below, we reproduce a small number of results from the cluster analysis field\cite{PH2003} which are relevant to the research presented in this paper.

Consider two timeseries labelled as $i$ and $j$. The dissimilarity measure between them $d_{ij}$ is a distance measure if it satisfies the triangle inequality
\begin{eqnarray}
d_{ij}+d_{jk} & \geq & d_{ik}
\end{eqnarray}
for timeseries $i$,$j$ and $k$ \cite{PH2003}. The following two conditions must also be met in order that the dissimilarity measure defines a meaningful distance
\begin{eqnarray}
d_{ij}=0 & \Longleftrightarrow & i=j\ , \\
d_{ij} & = & d_{ji}, \ \ \  \forall i,j\ .
\end{eqnarray}
Recall that timeseries $i$ is represented by a vector $\x_i$ and $x_{ik}$ is the $k$th component of this vector, where $k$ runs from 1 to $p$ where $p$ is the same for each timeseries. Let $w_k$ be a non-negative weight which is the same for the $k$th component of each timeseries.  Now consider the following three distance measures: the \emph{Euclidean Measure}, the \emph{Standardized Euclidean Measure} and the \emph{Correlation Measure}.
\begin{enumerate}
\item 
Euclidean Measure:
\begin{eqnarray}
d_{ij} &=& \left( \sum_{k=1}^p{w_k^2 (x_{ik}-x_{jk})^2} \right).
\end{eqnarray}
If one ignores the weighting terms $w_k$, this formula is simply the Euclidean Distance between two $p$-dimensional vectors. In fact, since there is no \emph{a priori} reason to weight the $k$th term differently to the $k$'th term, we shall not include it in any of the distance measures from this point onwards.

\item
Standardized Euclidean Measure:
\begin{eqnarray}
d_{ij} &=& \sqrt{\sum_{k=1}^{p}{\left(\frac{x_{ik}}{\left|\x_i\right|^2} - \frac{x_{jk}}{\left|\x_j\right|^2}  \right)}}, \\
d_{ij}^2 &=& \sum_{k=1}^{p}{\left( \frac{x_{ik}^2}{\left|\x_i\right|^2} + \frac{x_{jk}^2}{\left|\x_j\right|^2} - 2\frac{x_{ik}x_{jk}}{\left|\x_i\right|^2\left|\x_j\right|^2}\right)}.
\end{eqnarray}
But $\sum_{k=1}^{p}{x_{ik}^2}=\left| \x_i \right|^2$ and similarly for $j$. So,
\begin{eqnarray}
d_{ij}^2 &=& 2 \left( 1-\sum_{k=1}^{p}{\frac{x_{ik}x_{jk}}{\left|\x_i\right|^2\left|\x_j\right|^2}} \right).
\end{eqnarray}
In the case where the expected value is zero for each step of each timeseries\cite{ExpRetNote} then we have
\begin{eqnarray}
d_{ij} &=& \sqrt{2(1-\rho_{ij})}\ ,
\end{eqnarray}
where $\rho_{ij}$ is the statistical correlation between timeseries $i$ and $j$.

\item
Correlation Measure:
\begin{eqnarray}
d_{ij} &=& 1-\rho_{ij}\ .
\end{eqnarray}
\end{enumerate}
At first sight it might appear that the three distance measures above are quite different -- however strong relationships do exist between all of them\cite{PH2003}. Consequently it is the choice of clustering procedure which tends to dictate the quality of the resulting clustering which emerges, rather than the choice of distance measure.\cite{PH2003}

The clustering method used to form the MST is known in cluster analysis as the \emph{single-linkage} clustering procedure -- this in turn is sometimes called the \emph{nearest-neighbour} technique \cite{KR1990}. It is the simplest among an important group of clustering methods known collectively as Agglomerative Hierarchical Clustering Methods. The main problem with the single-linkage method (i.e. the MST) is that it has a tendency to link poorly clustered groups into `chains' by successively joining them through their nearest neighbours. Hence, one would expect the hierarchy produced by the MST to represent larger distances (i.e. anti-correlation) less reliably than it does smaller distances (i.e. high correlation). Since we are attempting to identify tightly-clustered groups in our data, this will not be a problem. However in other situations -- for example, if one were attempting to use an MST to identify poorly correlated or anti-correlated stocks for use in portfolio theory -- it may be preferable to use a more sophisticated clustering method.

\section{FX MARKET DATA}
\label{sect:data}  

We investigated the hourly, historical price-postings from HSBC Bank's database for nine currency pairs together with the price of Gold from 01/04/1993 to 12/30/1994 \cite{note8}.  We included Gold in the study because there are similarities in the way that it is traded, and in some respects it resembles a very volatile currency. The currency pairs under investigation are AUD/USD, GBP/USD, USD/CAD,	USD/CHF,	USD/JPY, GOLD/USD, USD/DEM, USD/NOK, USD/NZD, USD/SEK \cite{note2}. In the terminology used in FX markets \cite{note2}, USD/CAD is counter-intuitively the number of Canadian dollars (CAD) that can be purchased with one US dollar (USD). We must define precisely what we mean by hourly data, since prices are posted for different currency pairs at different times. We do not want to use average prices since we want the prices we are investigating to be prices at which we could have executed trades. Hence for hourly data, we use the last posted price within a given hour to represent the (hourly) price for the following hour. Before proceeding any further, we note that the $n$ stochastic variables which we will analyze correspond to currency {\em exchange} rates and hence measure the {\em relative} values of any two currencies. It is effectively meaningless to ask the {\em absolute} value of a given currency, since this can only ever be measured with respect to some other financial good. Thus each currency pair corresponds to a node in our network. We are concerned with the correlations between these currency exchange rates, each of which corresponds to an edge between two nodes. A given node does {\em not} correspond to a single currency. 

In common with all other real-world systems, the issue of what constitutes correct data is a complicated one. Most importantly, there are some subtle data-filtering (or so-called `data-cleaning') issues which need to be addressed. In our specific case, we are interested in calculating both the instantaneous and lagged correlations between exchange-rate returns. Hence it is neccessary to ensure that (a)
each time series has an equal number of posted prices; (b) the $k$'th posting for each currency pair corresponds, to as good an approximation as possible, to the price posted at the same timestep $t_k$ for all $k \in \{1, ... , p\}$. For some of the hourly timesteps, some currency pairs have missing data. The best way to deal with this is open to interpretation.\cite{compnote} Is the data missing simply because there has been no price change during that hour, or was there a fault in the data-recording system? Looking at the data, many of the missing points do seem to occur at times when one might expect the market to be illiquid. However, sometimes there are many consecutive missing data points --- even an entire day. This obviously reflects a fault in
the data recording system. To deal with such missing data we adopted the following protocol.  The FX market is at its most liquid between the hours of 08:00 and 16:00 GMT \cite{JJPS2004}. In an effort to eradicate the effect of `zero returns' due to a lack of liquidity in the market -- as opposed to the price genuinely not moving in consecutive trades -- we only used data from between these hours \cite{note5}. Then, if the missing data were for fewer than three consecutive hours, the missing prices were taken to be the value of the last quoted price. If the missing data were for three or more consecutive hours, then the data for those hours were omitted from the analysis. Since we must also ensure completeness of the data at each point, it is then necessary that the data for those hours are omitted from {\em all} currency pairs under investigation \cite{note6}. We believe that this procedure provides a sensible compromise between the conflicting demands of incorporating all relevant data, and yet avoiding the inclusion of spurious zero-returns which could significantly skew the data. Having cleaned up the exchange-rate data for each currency pair, whose associated price we will henceforth label as $P_i(t)$, we  turn this value into a financial return
\begin{eqnarray} 
r_i(t_n) & = & \ln \left( \frac{P_i(t_{n+1})}{P_i(t_n)}\right).
\end{eqnarray}

We need to be confident that our return distributions are stationary, since we wish to calculate correlations. A useful probe of stationarity is to calculate the autocorrelation for each time-series. A stationary time-series will have an autocorrelation function which rapidly decays to zero \cite{PR1998}, whereas a non-stationary time-series will have an autocorrelation function which decays to zero very slowly (if at all). The autocorrelation is defined as
\begin{eqnarray}
\rho_{i}(\tau) &=& \frac{\av{x_{i,t+\tau}x_{i,t}} -
\av{x_{i,t+\tau}}\av{x_{i,t}}}{\sigma_{i,\tau}\sigma_i}
\end{eqnarray} where $\av{...}$ indicates a time-average over the $p-\tau$ elements and $\sigma_{i,\tau}$, $\sigma_i$ are the sample standard deviations of the time-series $x_{i,t+\tau}$ and $x_{i,t}$ respectively. An analysis of the autocorrelation of both the price and return confirms that the returns are stationary whilst the prices themselves are not. Thus we can, with confidence, focus on correlations between different currency-pair returns.
 
There are a number of  other data-related issues which make the study of FX and equities fundamentally different. When producing the MST for the returns of the stock which make up the FTSE100 index, one calculates the returns
from the values of the price of the stock \emph {in the same currency} --- specifically, UK pounds (GBP). However, with FX data we are considering exchange rates between currency pairs. Thus should we consider GBP/USD or USD/GBP? And does it indeed make a difference which one we use? Since the correlation is constructed to be normalized and \emph{dimensionless}, one might be tempted to think that it does not matter since the value of the correlation will be the same and only the sign will be different. However, it is important when constructing the MST since there is an asymmetry between how positive and negative correlations are represented as distances. In particular, the MST picks out the smallest distances --- i.e. the highest correlation. A large negative correlation gives rise to a large distance between nodes. Thus a connection between two nodes will be missing from the tree even though it would be included if the other currency in the pair were used as the base currency.

Suppose there is a large negative correlation between the returns of the two currency pairs GBP/USD and USD/CHF \cite{note3}. Conversely, if we put them both with USD as the base currency, we get a large positive correlation between USD/GBP and USD/CHF. Thus our choice will give rise to a fundamentally different tree structure.
For this reason, we perform the analysis for all possible currency-pairs against each other. Since we are analysing ten currency pairs, this gives us eleven separate currencies and hence 110 possible currency pairs (and hence $n=110$ nodes). However, there are constraints on these timeseries and hence an intrinsic structure is imposed on the tree by the relationships between the timeseries. This is commonly known as the `triangle effect'. Consider the three exchange-rates USD/CHF, GBP/USD and GBP/CHF. The $n$th element of the timeseries for GBP/CHF is simply the product of the $n$th elements of USD/CHF and GBP/USD. This simple relationship between the timeseries gives rise to some relationships between the correlations. More generally, with three time-series $P_1(t)$,$P_2(t)$,$P_3(t)$ such that $P_3(t) = P_1(t)P_2(t)$, there exist relationships between the correlations and variances of the returns. If we define
the returns $r_i$ such that $r_i = \ln P_i$ for all $i$, then we have:
\begin{eqnarray} 
r_3 = r_1 + r_2\ .
\end{eqnarray} Thus
\begin{eqnarray}
\Var(r_3) &=& \Var (r_1+r_2)\\
 &=& \E((r_1+r_2)^2) - (\E(r_1+r_2))^2
\end{eqnarray} 
For currency pairs, it is valid to assume that the expected value of the return is zero \cite{note4}. Hence this expression simplifies to
\begin{eqnarray}
\sigma_3^2 &=& \E(r_1^2+r_2^2+2r_1r_2)\\
 &=& \sigma_1^2 + \sigma_2^2 + 2\Cov(r_1,r_2)\\
 &=& \sigma_1^2 + \sigma_2^2 + 2\sigma_1\sigma_2\rho_{12}
\label{good}
\end{eqnarray} where $\sigma_1$, $\sigma_2$, $\sigma_3$ are the variances of the returns $r_1(t)$, $r_2(t)$, $r_3(t)$ while $\rho_{12}$ is the correlation between the returns $r_1(t)$ and $r_2(t)$. Finally we obtain
\begin{eqnarray}
\rho_{12} &=& \frac{\sigma_3^2 - (\sigma_1^2+\sigma_2^2)}{2\sigma_1\sigma_2}
\label{Triangle}
\end{eqnarray}
Hence there is a structure forced upon the market by the triangle effect. This is not a problem since all the cross-rates we include in the tree \emph{do} exist and the correlations calculated are the true correlations between the returns. Even though the values of these correlations have some relationships between them, they should be included in the tree since it is precisely this market structure that we are attempting to identify. We will, however, need to confirm that this structure which is being imposed on the market is not dominating our results. 

We performed a number of tests to ensure the validity of our data. One simple check performed was to calculate the minimum and maximum return for each currency pair. For example, if the rate for USD/JPY was entered as 1.738 instead of 173.8 then this would give rise to returns of approximately $\pm 1$. As a result of this check, we could confirm that there were no such errors in our dataset. We then drew scatter plots of the returns against time, plus histograms of the return distribution in order to check  that there were no unusual points on the graph. The next check that we performed is slightly more subtle. The correlation between two variables is related to the gradient of the regression line between the variables \cite{PR1998}. However, this gradient is very susceptible to outliers so we need to ensure that our data does not contain such outlying points. To check this, one can plot scatter-graphs of returns from different currency pairs against each other to confirm that there are no points sufficiently outlying as to justify deletion.

\section{MST FOR CURRENCIES}
\label{sect:realmst}
The Minimum Spanning Tree approach was generalized in Ref. \cite{KKK2003} by considering a directed graph. Lagged correlations were investigated in an attempt to determine whether the movement of one stock price `preceded' the movement in another stock price. We now investigate whether this approach yields useful results here. First we should define what we mean by lagged correlation. If we have two time-series, $x_i(t)$ and $x_j(t)$ where both time-series contain $p$ elements, the $\tau$-lagged correlation is given by
\begin{eqnarray} 
\rho_{ij}(\tau) &=& \frac{\langle x_{i,t+\tau}x_{j,t} \rangle - \langle x_{i,t+\tau}
\rangle \langle x_{j,t} \rangle}{\sigma_{i,\tau} \sigma_j}
\end{eqnarray} 
   \begin{figure}[t]
   \begin{center}
   \begin{tabular}{c}
   \includegraphics[height=7cm]{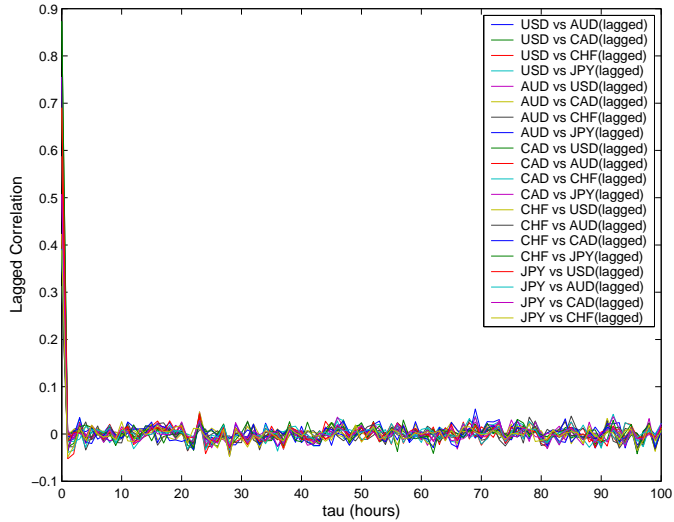}
   \end{tabular}
   \end{center}
   \caption[LagGBP] 
   { \label{LagGBP}
Lagged correlation between different currency pairs when GBP is the base currency. As explained in the text, AUD vs USD (lagged) refers to the lagged correlation between GBP/AUD (at time $t+\tau$) and GBP/USD (at time $t$)}
   \end{figure} 

\noindent where $\langle$...$\rangle$ indicates a time-average over the $p-\tau$ elements and $\sigma_{i,\tau}$, $\sigma_j$ are the sample standard deviations of the time-series $x_{i,t+\tau}$ and $x_{j,t}$ respectively. Note that the autocorrelation $\rho_{i}(\tau)$ which was defined earlier, is simply the special case of $\rho_{ij}(\tau)$ where $i=j$. Armed with this definition, we can now look at our data to see whether there are any significant lagged correlations between returns of different currency pairs. Figure \ref{LagGBP} shows the lagged correlation between the returns of each pair of currencies when the prices of those currencies are given with GBP as the base currency. In the figure, AUD vs USD (lagged) refers to the lagged correlation between GBP/AUD (at time $t+\tau$) and GBP/USD (at time $t$). The results in this figure are representative of the results from all currency pairs included in our study. Figure \ref{LagGBP} clearly shows that any lagged correlations which might exist, will only occur over a timescale smaller than one hour. Hence we will not consider lagged correlations any further in the present paper. This lack of any noticeable lagged correlations implies that the FX market is very efficient, which in itself should not come as a surprise --- after all, the FX market is approximately 200 times as liquid as the equities market \cite{PEV2003}.

   \begin{figure}
   \begin{center}
   \begin{tabular}{c}
   \includegraphics[width=.75\textwidth]{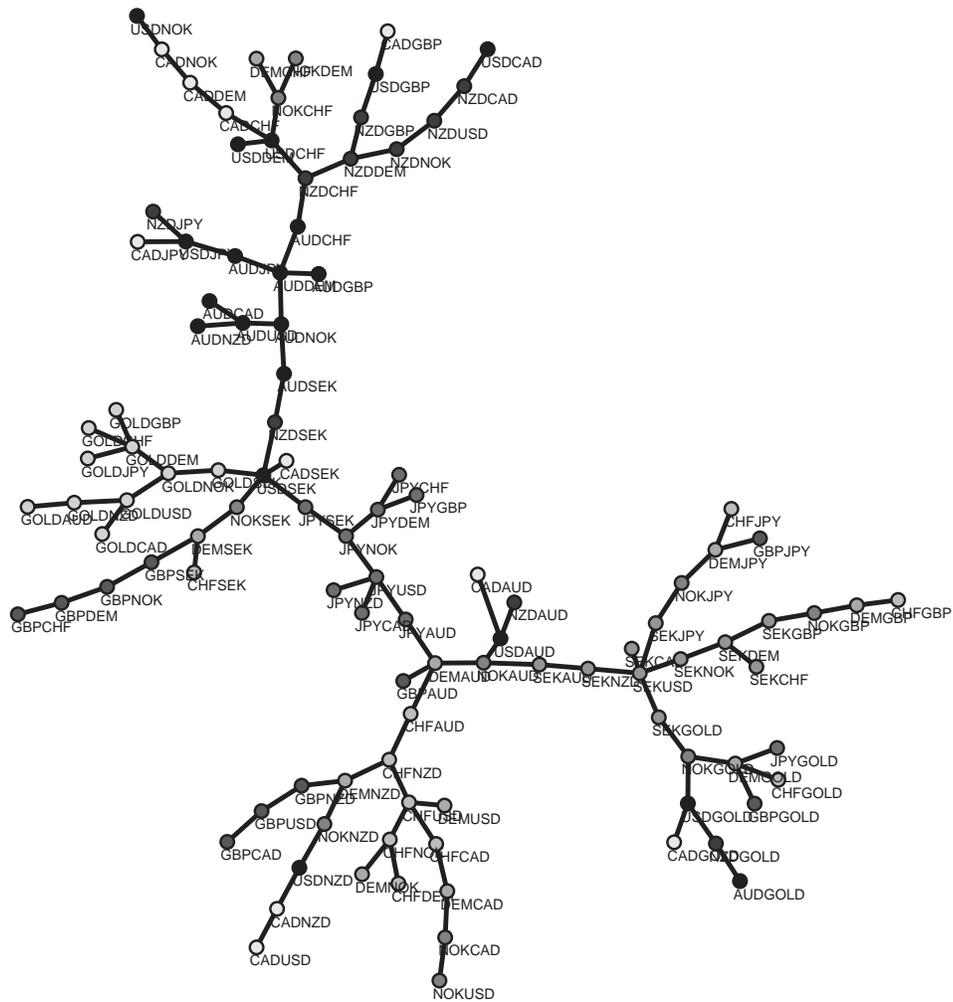}
   \end{tabular}
   \end{center}
   \caption[All9394] 
   { \label{All9394}
The Minimum Spanning Tree (MST) representing the correlations
between all hourly cross-rate returns from the years 1993 and 1994. Created using
the well-known Pajek software.}
   \end{figure} 

{C}reating all the possible cross-rates from the 11 currency pairs gives rise to a total of $n=110$ different time-series. It is here that the approach of constructing the MST comes into its own, since 110 different currencies yields an enormous correlation matrix containing 5995 separate elements. This is far too much information to allow any practical analysis by eye. However, as can be seen from Figure \ref{All9394}, the hourly FX tree is quite easy to look at. Rather than a mass of numbers,  we now have a graphical representation of the complex system in
which the structure of the system is visible.

\begin{figure}
\begin{center}
\begin{tabular}{c}
\includegraphics[width=.75\textwidth]{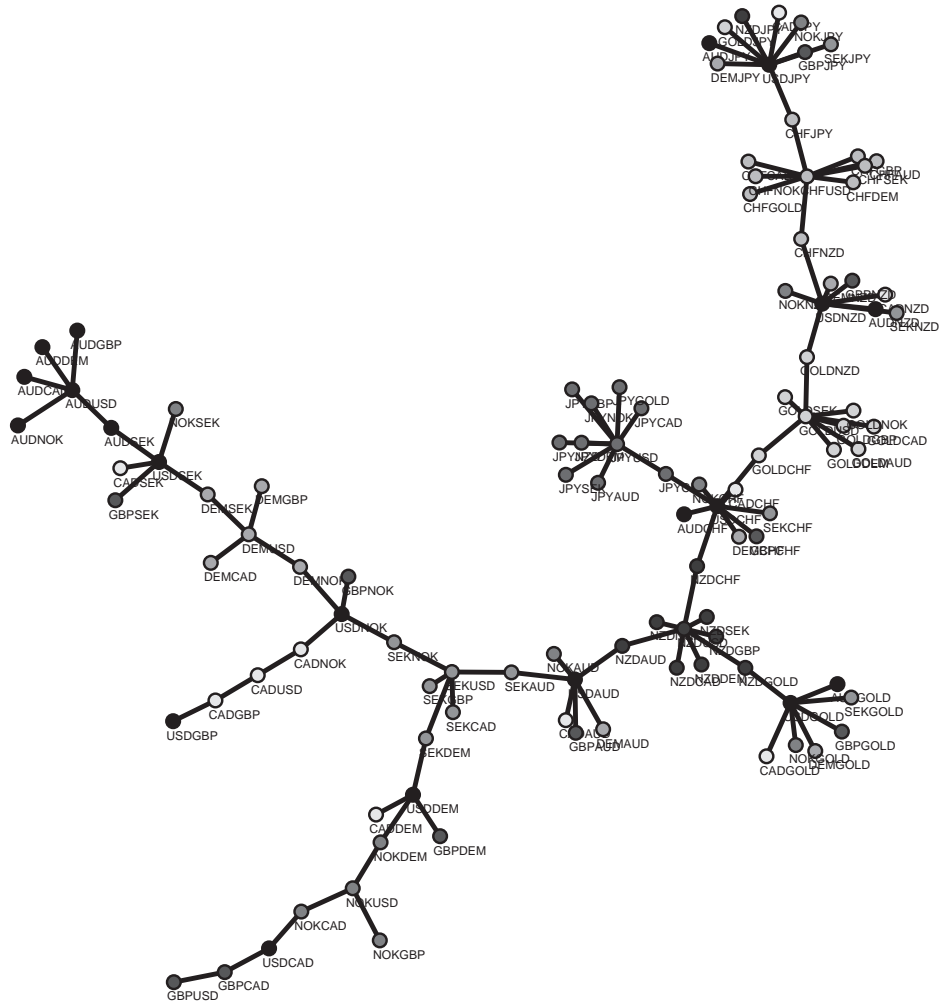}
\end{tabular}
\end{center}
\caption[jumbled]
{ \label{jumbled}
The Minimum Spanning Tree (MST) formed from randomized data for the USD prices.
This shows only the structure imposed on the tree by the triangle effect.}
\end{figure}

Before analyzing the tree in detail, we consider first what effect the constraints of Eq. (12)  (the `triangle effect') will have on the shape of the tree. This can be checked by generating the MST for price-series for the currencies in USD which have been randomized before the cross-rates were formed. This process gives prices for the various currencies in USD which are random, and will hence have negligible correlation between their returns. The resulting MST is shown in Fig. \ref{jumbled}. As can be seen from the figures, the MST for the randomized data is very different in character from the true tree in Figure \ref{All9394}. At first glance it might appear that some aspects are similar --- currencies show some clustering in both cases. However, in the tree of real cross-rates there are currency-clusters forming about any node, whereas in Figure
\ref{jumbled} there are only clusters centred on the USD node. This is not surprising: what do the `CHF/everything' rates all have in common in the case of random prices other than the CHF/USD rate?  The best way to interpret Fig. \ref{jumbled} is that we have a tree of USD nodes which are spaced out since their returns are poorly correlated, and around these nodes we have clusters of other nodes which have the same base currency and which are effectively the information from the USD node plus noise. 

\begin{figure}
   \begin{center}
   \begin{tabular}{c}
   \includegraphics[height=7cm]{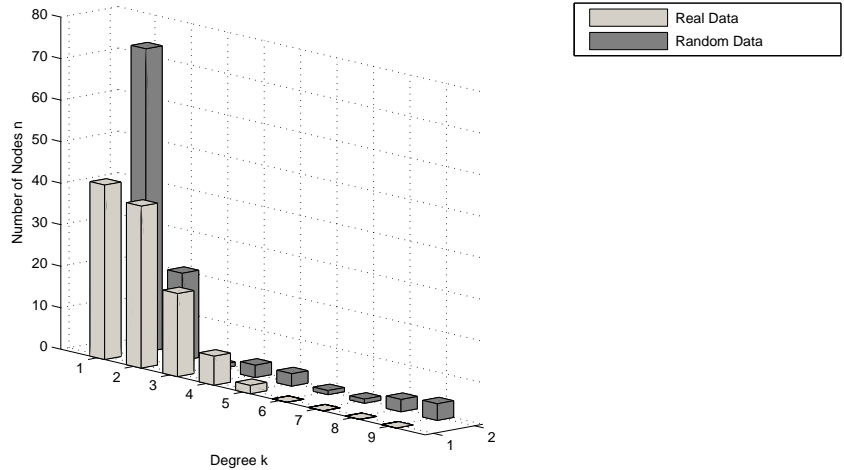}
   \end{tabular}
   \end{center}
   \caption[degreedist] 
   { \label{degreedist}
Comparison of the degree distributions for the trees shown in Figure
\ref{All9394} (Real Data) and Figure \ref{jumbled} (Randomized Data).}
  \end{figure}

This exercise has shown us that the MST results are not dominated by the triangle effect. In an effort to show this in a more quantitative way, we also investigated the proportion of links that are present in both trees. Less than one third of the edges in Fig. \ref{All9394} are present in Fig. \ref{jumbled}.
We also compared the degree distribution of the tree from the random price series with that of the tree from real price data. This is shown in Fig. \ref{degreedist} and again demonstrates the differences between the two trees.

Having produced the MST, we now discuss its interpretation. The tree contains nodes, each of which represents a particular currency pair. For the reasons explained earlier, currency pairs are quoted both ways round: USD/JPY appears with USD as the base currency, as is normal market convention, but so does JPY/USD. This gives all currencies the chance to stand out as a cluster, as will be seen shortly. Broadly speaking, each node is linked to the nodes representing the currency-pairs to which it is most closely correlated. The observation that certain currency-pairs cluster together means that they have been moving together consistently over the monitored period.
The most interesting feature of Figure \ref{All9394} is the clustering of nodes which have the same base currency. For example, one can see a cluster of 9 AUD nodes. This observation demonstrates that over this two year period, the Australian Dollar has been moving systematically against a range of other currencies during this time. To use the prevailing industry term, the AUD is `in play'. The same is also true for the SEK, JPY and GOLD clusters.

   \begin{figure}
   \begin{center}
   \begin{tabular}{c}
   \includegraphics[height=7cm]{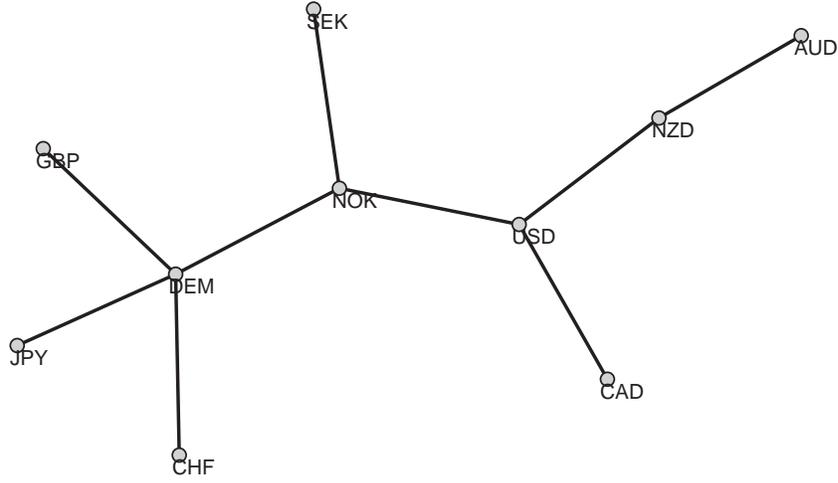}
   \end{tabular}
   \end{center}
   \caption[Gold] 
   { \label{Gold}
The cluster of GOLD exchange-rates from Fig. \ref{All9394}.}
  \end{figure}

The cluster of Gold exchange rates links currencies in a sensible way, which is an encouraging sign. This cluster is re-drawn in Figure \ref{Gold}. It can be seen that the nodes in this cluster are grouped in an economically meaningful way: remarkably, there is a geographical linking of exchange-rates. The Australisian nodes, AUD and NZD, are linked, as are the American ones (USD and CAD). The Skandinavian currencies, SEK and NOK, are also linked. Finally, there is a European cluster of GBP, CHF and EUR. 

Given that it is possible to identify clusters of currencies, we would like to quantify how clustered they are. This can be done by finding the level one has to partition the hierarchical tree associated with the MST \cite{RTV1986} to get all the nodes with, for example, GBP as the base currency into the same cluster. This results in a \emph{self-clustering distance} for each currency. The smaller this distance is, the more tightly all the nodes for that currency are clustered. An alternative way to think of this is as the maximum ultrametric distance between any two nodes for that currency. 

\begin{figure}[h]
\begin{center}
\includegraphics[width=.47\textwidth]{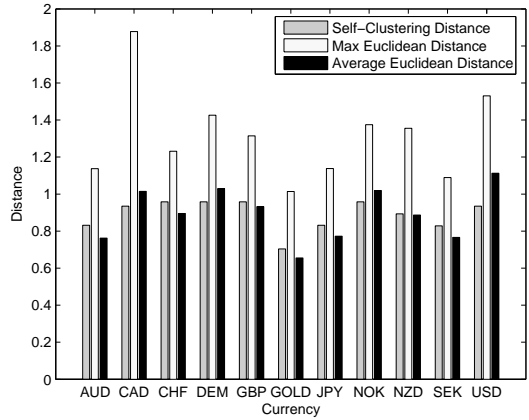}
\caption{Comparison of the results from the MST with those from the original correlation matrix}
\label{Compare}
\end{center}
\end{figure}

We are now in a position to compare the results produced by the MST with those from the original distance matrix itself. In particular, we will compare the self-clustering distance for each currency with the maximum Euclidean Distance between any two nodes which contain that base currency, and also with the average Euclidean Distance between all nodes which contain that base currency. These results are shown in Figure \ref{Compare}. It can be seen that the agreement is very good. Not only does the MST rank the clusters in the same way as the original distance matrix, but it also gives results which agree better with the average distance than with the maximum Euclidean distance. Hence the results for the MST and the original distance matrix are not only in agreement, they are also robust with respect to a single large edge being contained between two nodes in the cluster. As mentioned earlier, the MST has a remarkable advantage over standard network representations since it only requires $n-1$ connections.

\section{TEMPORAL EVOLUTION OF MST FOR CURRENCIES}
\label{sect:tempevol}
The single-step survival ratio of the edges (i.e. connections) is defined as 
\begin{eqnarray}
\sigma_{\delta t} &=& \frac{\left| E_t\cap E_{t+\delta t} \right|}{\left| E
\right|}
\end{eqnarray} where $E_t$ and $E_{t+\delta t}$ represent the set of edges (i.e. connections) present in the trees formed from a dataset of length $T$=1000 hours \cite{note7} beginning at times $t$ and $t+\delta t$ respectively, in order to see how it depends on the value chosen for $\delta t$. This ratio tends to one as $\delta t$ approaches 0, hence the topology of the MST is stable.
Based on this, Onnela \cite{JPO2002} defined the $k$ multi-step survival ratio as
\begin{eqnarray}
\sigma_{\delta t,k} &=& \frac{\left| E_t \cap E_{t+\delta t} \cap \ldots \cap
E_{t+k\delta t}  \right|}{\left| E \right|}
\label{min}
\end{eqnarray} Thus if a link disappears for only one of the trees in the time $t$ to $t+\delta t$ and then comes back, it is not counted in this survival ratio. This seems a somewhat over-restrictive definition since it could underestimate the survival. We will therefore also consider the more generous definition
\begin{eqnarray}
\sigma_{\delta t,k} &=& \frac{\left| E_t \cap E_{t+k\delta t} \right|}{\left| E
\right|}\ \ .
\label{max}
\end{eqnarray} 
This quantity will, for large values of $k$, include cases where the links disappear and then come back several timesteps later. It therefore tends to overestimate the survival since a
reappearance after such a long gap is more likely to be caused by a changing structure than by a brief, insignificant fluctuation.

   \begin{figure}
   \begin{center}
   \begin{tabular}{c}
   \includegraphics[height=7cm]{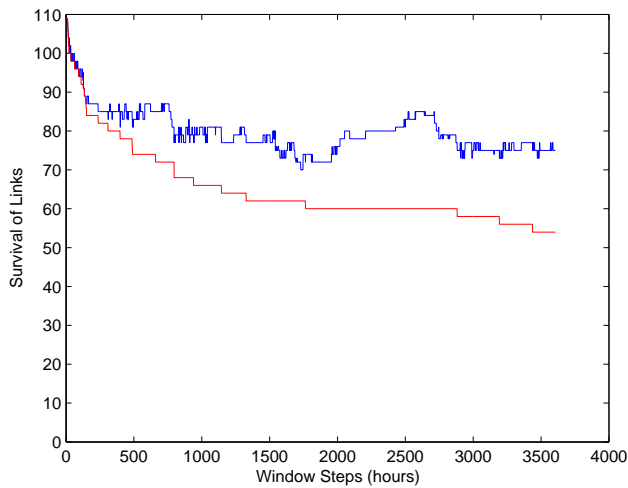}
   \end{tabular}
   \end{center}
   \caption[MultiStep] 
   { \label{MultiStep}
Multi-step survival ratio of the FX tree's connections, as a function of
time. The graph shows the two definitions described in the text, which tend to
overestimate (upper line) and underestimate (lower line) the survival effect.}
  \end{figure}

Figure \ref{MultiStep} shows both definitions, and  uses a time-window of length $T=1000$ hours and a time-step $\delta t=1$ hour \cite{note7}. It can be seen from the figure that the two lines form a `corridor' for the multi-step survival ratio. This is because the over-restrictive definition of Eq. (\ref{min}) under-estimates the survival and the over-generous definition of Eq. (\ref{max}) over-estimates the result. It is particularly noteworthy that even with the over-restrictive definition of Eq. (\ref{min}), the survival of links after the end of two years is only just below fifty percent (i.e. 54/109). In other words, there are strong correlations existing between exchange-rate returns that are extremely long-lived.

\section{CONCLUSIONS AND FUTURE DIRECTIONS}
\label{sect:conclus}
We have provided a detailed analysis of the correlation networks, and in particular the Minimal Spanning Trees, associated with empirical data obtainined from the Foreign Exchange (FX) currency markets. This analysis has highlighted various data-related features which make this study quite distinct from earlier work on equities. 
There is a clear difference between the resulting currency trees formed from real markets and those formed from randomized data. Not only does the tree look different visually, but also the degree distributions of the two trees are markedly different.  For the trees from real markets, there is a clear regional clustering. We have also investigated the time-dependence of the trees. Even though the market structure does change rapidly enough to identify changes in which currency-pairs are clustering together, there are links in the tree which last over the entire two year period. This shows that there is a certain robust dynamics to the FX markets. At the same time, our analysis of the dynamical evolution of the MST shows that there is an effective `ecology of clusters': in other words, clusters can survive for finite periods of time during which time they may evolve in some identifiable way, before eventually dissipating or `dying'. This supports the view that the FX market is a continually evolving ecology -- one could even say that it is `alive' in a dynamical sense. Future work will look at  identifying which currencies are actively in play and are effectively dominating the FX market. Furthermore, we shall also be reporting studies on the extent to which external news will `shake' the FX tree by supplying occasional kicks. Of particular interest is whether particular clusters have increased robustness over others, or not. In addition, we shall be investigating how the observed tree structure and its temporal evolution, depend on the frequency of the data used.

Future work will also look toward identifying candidate multi-agent models of artificial markets\cite{book} -- and in particular the set of possible population compositions for the multi-agent game itself -- which generate MSTs which are consistent with that observed in the FX market. We suspect that MSTs could be used to guide the selection of populations in such a multi-agent model, in order to produce timeseries in accordance with the respective statistical stylized facts of the individual exchange-rate timeseries, and indeed even to develop a cross-market predictive tool. We note that a simplified, prototype version of such a multi-agent model was presented several years ago \cite{ejpb,details} and was shown to have a surprisingly high predictive power for the Dollar-Yen exchange-rate.

\acknowledgments    
We thank David Smith, Nachi Gupta and Kurt Mitman for discussions.


\begin{thebibliography}{1}

\bibitem{econo} See www.unifr.ch/econophysics for an up-to-date listing of all
network papers, preprints and books. 

\bibitem{nets1} D.J. Watts and S.H. Strogatz, Nature {\bf 393}, 440 (1998).

\bibitem{nets2} D.S.
Callaway, M.E.J.  Newman, S.H.  Strogatz, and D.J. Watts, Phys. Rev. Lett. {\bf
85}, 5468 (2000).

\bibitem{DM2003} S.N. Dorogovtsev and J.F.F. Mendes, {\em Evolution of Networks:
From Biological Nets to the Internet and WWW}, Oxford University Press, Oxford,
2003. 

\bibitem{book} For example, see N.F. Johnson, P. Jefferies and P.M. Hui, {\em Financial Market Complexity}, Oxford University, New York, 2003.


\bibitem{taylor} F. Taylor, {\em Mastering FX and Currency
Options}, Financial Times Press, London, 1997.

\bibitem{PEV2003} P. Evans, {\em Introduction to Treasury and Capital Markets},  
HSBC Bank Training Manual, 2003.


\bibitem{MRN1999} R.N. Mantegna, 
Eur. Phys. Jour. B {\bf 11}, 193 (1999).

\bibitem{MS2000} R.N. Mantegna and H.E. Stanley, {\em An Introduction to
Econophysics: Correlations and Complexity in Finance}, Cambridge University Press,
Cambridge, 2000.

\bibitem{JPO2002} J.-P. Onnela, {\em Taxonomy of Financial Assets}, M.Sc.
Thesis, Department of  Electrical and Communications Engineering, Helsinki University
of Technology (2002).

\bibitem{OCKK2003b} J.-P. Onnela, A. Chakraborti,  K. Kaski, and J. Kertesz, 
Phys. Scrip. T {\bf 106}, 48 (2003). 

\bibitem{OCKK2003a} J.-P. Onnela, A. Chakraborti, K. Kaski, and J. Kertesz,
Physica A {\bf 324}, 247 (2003). 

\bibitem{OCKK2003c}  J.-P. Onnela, A. Chakraborti, K. Kaski, and J. Kertesz,
Phys. Rev. E  {\bf 68}, 056110 (2003).

\bibitem{GB2003} G. Bonanno \emph{et at.}, Phys. Rev. E {\bf 68}, 046130 (2003).

\bibitem{GB2004} G. Bonanno \emph{et at.}, Eur. Phys. J. B {\bf 38}, 363 371 (2004).

\bibitem{note9} We assume that the correlation structure is specified by the correlation coefficients.


\bibitem{RTV1986} R. Rammel and G. Toulouse, and M.A. Virasoro, Rev. Mod.
Phys. {\bf 58}, 765 (1986).

\bibitem{note1} For a discussion as
to why this function is used as the definition of distance and not, say,
$d_{ij}=1-\rho_{ij}$, see Chapter 13 of Ref. 9. We show in Section 3 that this formula simply approximates to the Standardized Euclidean Distance Measure between two time series ${\underline x}_i$ and ${\underline x}_j$. The reason this formula appears so counter-intuitive is because the result is given in terms of the correlation between the two time-series rather than
as a sum over the elements of the series.

\bibitem{BBo1979} B. Bollobas, {\em Graph Theory, An Introductory Course}, Springer-Verlag, New York, 1979.

\bibitem{CLR1990} T.H. Cormen, C.E. Leiserson, and R.L. Rivest, {\em
Introduction to Algorithms}, MIT Press, Cambridge MA, 1990. 

\bibitem{PH2003} P. Hirst, \emph{Cluster Analysis of Financial Securities}, M.Sc. Thesis, Oxford Centre for Industrial and Applied Mathematics, Oxford University (2003)

\bibitem{ExpRetNote} We note that for the currency pairs we investigated the interest-rate differentials are small and the time-period of one hour is tiny. 

\bibitem{KR1990} L. Kaufman and P.J. Rousseeuw, \emph{Finding Groups in Data. An Introduction to Cluster Analysis}, Wiley-Interscience, New York, 1990.

\bibitem{note8} This is over 12000 datapoints. 

\bibitem{note2} The acronyms are standard ones in the FX market, where AUD means
Australian dollar, GBP is UK pound, CAD is Canadian dollar etc. The exception is
GOLD/USD which is quoted as XAU/USD in the markets -- however for clarity, we have
replaced the rather obscure symbol XAU with GOLD. For more information on terminology
and functioning of the FX markets, see Ref. 6.

\bibitem{compnote} There is an extensive literature on data completion, which is summarized nicely in Ref. 27. The data completion method we use is known as Absolute Value Extrapolation and has the benefit of being one of the simplest.

\bibitem{RR2003} R. Rossmanith, \emph{Completion of Market Data}, M.Sc. Thesis, Oxford Centre for Industrial and Applied Mathematics  (2003)

\bibitem{JJPS2004} M.H. Jensen, A. Johansen, F. Petroni, and I. Simonsen, 
Inverse Statistics in the Foreign Exchange Market,  Physica A  {\bf 340} 678 (2004).

\bibitem{note5} If there were no missing data points, this would give rise to a dataset of 4680 data points.

\bibitem{note6} This gave rise to the final dataset which contained 4608 datapoints. Therefore, the data cleaning procedure removed less than 2 percent of the 4680 possible datapoints.

\bibitem{PR1998} R.S. Pindyck, and D.L. Rubinfeld, {\em Econometric Models and
Economic Forecasts}, McGraw-Hill, New York, 1998.

\bibitem{note3} Even though quoting the two exchange-rates in this form seems slightly idiosyncratic, it is actually how the rates are quoted on the market. Most currencies are quoted with USD as the base currency. However GBP/USD is quoted in this unusual form because
historically GBP was not based on a decimal system.

\bibitem{note4} We note that the interest-rate differentials are small and the time-period of one hour is tiny. We also note that when price-changes are small\cite{MS2000}, it makes little difference in terms of the overall behavior as to which precise definition of returns one uses. 

\bibitem{KKK2003} L. Kullmann, J. Kertesz, and K. Kaski, 
Phys. Rev. E {\bf 66} 026125 (2002).

\bibitem{note7} $T=$1000 hourly datapoints corresponds to just under half a year of data.

\bibitem{ejpb} N.F. Johnson, D. Lamper, P. Jefferies, M.L. Hart, S. Howison, Physica A {\bf 299}, 222 (2001); P. Jefferies, N.F. Johnson, M. Hart, P.M. Hui, Eur. J. Phys. B {\bf 20}, 493 (2001).

\bibitem{details} Details of the methods we developed at Oxford University for building multi-agent models with some predictive power, are given in D. Lamper's D.Phil. Thesis, Mathematics Institute, Oxford University, 2002. These details include the use of Kalman Filters to represent the weights on different binary strategies, an appropriate choice of memory length $m$ for the agents (e.g. $m=1$) and details about how to combine the predictions from candidate games with different parameters (e.g. a majority voting procedure, winner-takes-all, or some form of optimization over a population of candidate games). Full details of our more recent and sophisticated 
multi-agent predictive models, 
will be given by N. Gupta, D.M.D. Smith, K. Mitman, M. McDonald, O. Suleman, C. Gou, R. Hauser and N.F. Johnson, in preparation (2005).


\end{thebibliography}
\end{document}